\documentclass[12pt,preprint]{aastex}
\shorttitle{The pulsation spectrum of VX Hydrae}
\shortauthors{Templeton et al.}

\begin{document}
\title{The pulsation spectrum of VX Hydrae}
\author{M.R. Templeton\altaffilmark{1,2}, G.~Samolyk\altaffilmark{1,3},
S.~Dvorak\altaffilmark{1}, R.~Poklar\altaffilmark{1}, 
N.~Butterworth\altaffilmark{1}, H.~Gerner\altaffilmark{1}
}
\altaffiltext{1}{AAVSO, 49 Bay State Road, Cambridge, MA 02138, United States}
\altaffiltext{2}{email: matthewt@aavso.org}
\altaffiltext{3}{email: gsamolyk@wi.rr.com}

\begin{abstract}
We present the results of a two-year, multisite observing campaign
investigating the high-amplitude $\delta$ Scuti star VX Hydrae
during the 2006 and 2007 observing seasons.  The final data
set consists of nearly 8500 $V$-band observations spanning HJD 2453763.6
to 2454212.7 (2006 January 28 to 2007 April 22).  Separate analyses of
the two individual seasons of data yield 25 confidently-detected
frequencies common to both data sets, of which two are pulsation modes,
and the remaining 23 are Fourier harmonics or beat frequencies of these two 
modes.  The 2006 data set had five additional frequencies with amplitudes less 
than 1.5 mmag, and the 2007 data had one additional frequency.  Analysis
of the full 2006-2007 data set yields 22 of the 25 frequencies found in
the individual seasons of data.  There are no significant peaks in the
spectrum other than these between 0 and 60 c/d.  The frequencies of the
two main pulsation modes derived from the 2006 and 2007 observing
seasons individually do not differ at the level of $3\sigma$, and thus
we find no conclusive evidence for period change over the span of these
observations.  However, the amplitude of $f_{1} = 5.7898$ c/d
changed significantly between the two seasons, while the amplitude of
$f_{0} = 4.4765$ c/d remained constant; amplitudes of the Fourier harmonics
and beat frequencies of $f_{1}$ also changed.  Similar behavior was
seen in the 1950s, and it is clear that VX Hydrae undergoes significant 
amplitude changes over time.
\end{abstract}


\section{Introduction}

The high-amplitude $\delta$ Scuti star VX Hya was discovered by
\citet{Hoffmeister1931}, and subsequent observational work by
\citet{Lause1938} made clear that VX Hya was a complex pulsator.  In the
1950's and 1960's \citet{Fitch1966} conducted long-term $UBV$
photometric monitoring of this star and found that changes occurred in the
pulsation behavior on timescales of a year or less such that a single
set of Fourier coefficients could not model the star over the long term.
\citet{Fitch1966} modeled VX Hya over these short time
scales using a Fourier series consisting of two pulsation frequencies,
the complete set of Fourier harmonics and beat frequencies through
fourth-order, and a long period of about 4400 days.  The physical origin
of the long 4400-day period was not explained; it was used primarily to
fit the data mathematically, and thus its reality is not certain.  The
underlying cause for the long-term complexity of VX Hya has remained
unexplained to the present day.

The limited time-series observations published subsequent to the work of
Fitch have not clarified the long-term pulsation changes, although both
photometry and theoretical modeling seem to make clear that VX Hya is an
otherwise normal $\delta$ Scuti star.  Str\"{o}mgren photometry
\citep{Breger1977} indicated that the star has near-solar abundances,
and a mass over 2M$_\odot$.  The observed fundamental period and ratio
of first overtone and fundamental periods agrees with stellar
models computed using modern opacities and abundances;
\citet{Templeton2000} suggests the star is slightly more massive than 2.3-2.4
M$_\odot$ with solar-like abundances.  The only
apparent peculiarity is the long-term change in its pulsations, and
long-term changes are no longer so peculiar among $\delta$ Scuti stars. 
Both period and amplitude changes are observed in many other $\delta$ Scuti
stars \citep{Handler2000,Arentoft2001,Breger03,BP2006}.  However, the 
physical causes behind them remain unknown. 

It is also unclear how these changes manifest themselves over time,
primarily because the observational records for nearly all $\delta$
Scuti stars are too sparse over the long term to give anything more than
a snapshot of how a star might be behaving at that one epoch in time. 
For example, the existing data for most stars cannot show whether long-term
period or amplitude changes are cyclical.
A few stars, primarily high-amplitude stars whose
time-of-maximum data are regularly collected, have long-term
observational records, and period changes for those are reasonably well
recorded.  However, time-of-maximum data for multiperiodic stars are
much harder to interpret \citep{Zhou2001,KDR06}.  For these stars, a direct
analysis of time-series photometry can provide a clearer picture of the
long-term behavior, as well as provide a means to uncover as yet
undetected variability.

In 2006 and 2007 a group of observers from the {\em American Association
of Variable Star Observers} (AAVSO) undertook a campaign to intensively
observe VX Hydrae's complex light curve.  Five observers obtained a
combined 8498 $V$-band observations of this star during the two
observing seasons, and the resulting light curves provide the best
seasonal coverage of this star to date.  This light curve
is an excellent data set from which to compute the current pulsation
spectrum of VX Hydrae, and it provides a new baseline for comparison with
both past and future data sets.  In this work, we present a
comprehensive analysis of these data. In Section 2 of this paper, we
describe the observational data obtained and their reduction to a common
system.  In Section 3, we give full details of the time-series analysis
and derivation of the Fourier spectrum and its changes from 2006 to
2007.  In Section 4, we discuss the results of the analysis, and put
this result in context of what is already known about VX Hydrae.

\section{Data}

The data set presented in this paper was collected by the five observers
at five different locations: Florida, Wisconsin, Arizona, and west Texas
in the United States, and Queensland in Australia.  The observers all
used telescopes with apertures of 0.2 to 0.25-meters, SBIG ST7 or
ST9 series cameras, and standard Johnson (Bessell) $V$ filters; see
Table \ref{table1} for details.  Approximately 8800 $V$-band data points
were obtained using TYC 5482-1347-1 (RA 09:45:29.56, Dec -11:58:45.8;
V=11.577, (B-V)=0.967) as the comparison star.  Following the removal of
discrepant and overlapping data, the final light curve contains 8498
differential measurements taken over two seasons from HJD 2453763.63564
to 2454212.71478.  These data were not extinction corrected or
transformed to a standard system, but have all been placed on a common
zero point.  The lack of extinction correction may introduce a small
nightly trend in the data, but absolute calibration is not required
because any constant terms are removed during Fourier transformation. 
The photometric error per point varies by observer and by night, but
typical values are on the order of 10 mmag or less. 

\begin{figure}
\begin{center}
\includegraphics[]{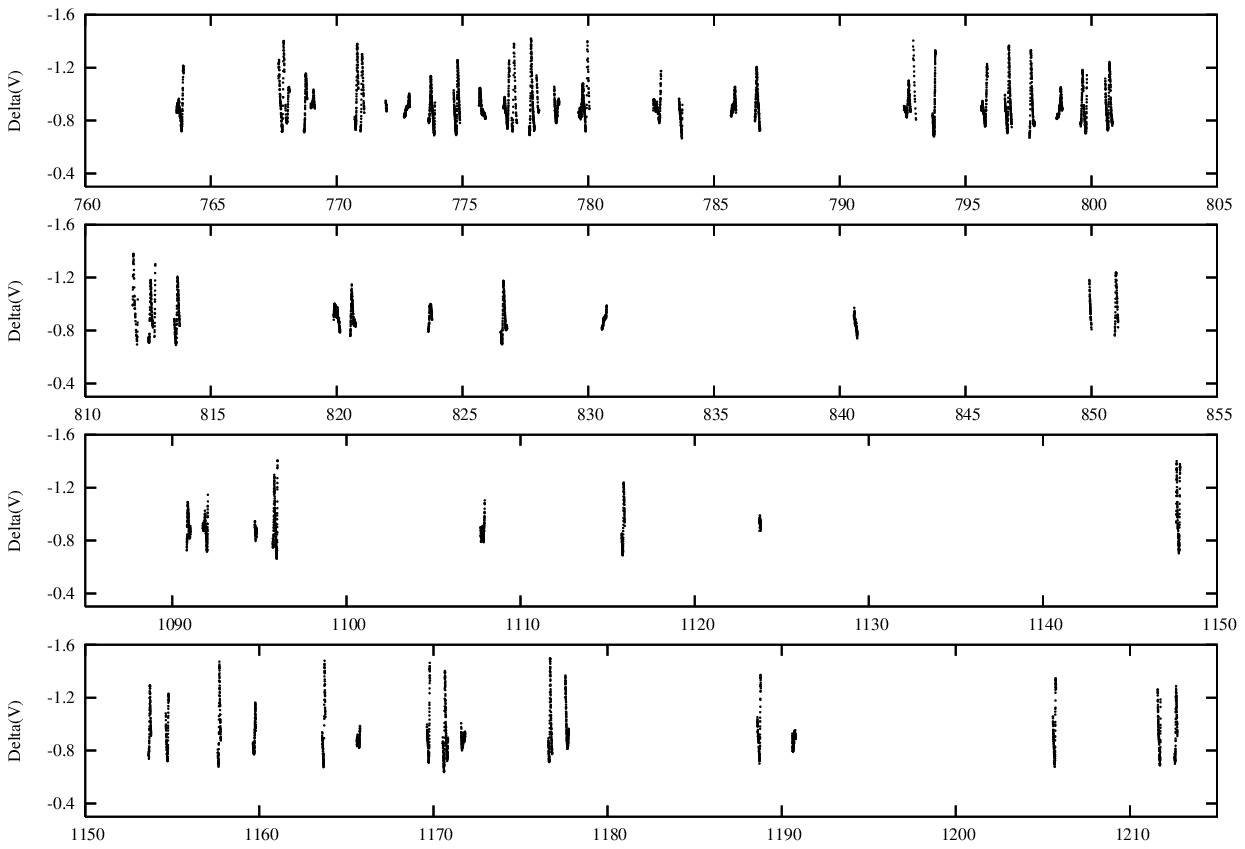}
\caption{The 2006-2007 light curve of VX Hydrae.  Data were normalized to the 
photometric mean light prior to Fourier analysis.  Times are days of HJD relative to HJD 2453000.}
\label{fig01}
\end{center}
\end{figure}

\begin{figure}
\begin{center}
\includegraphics[]{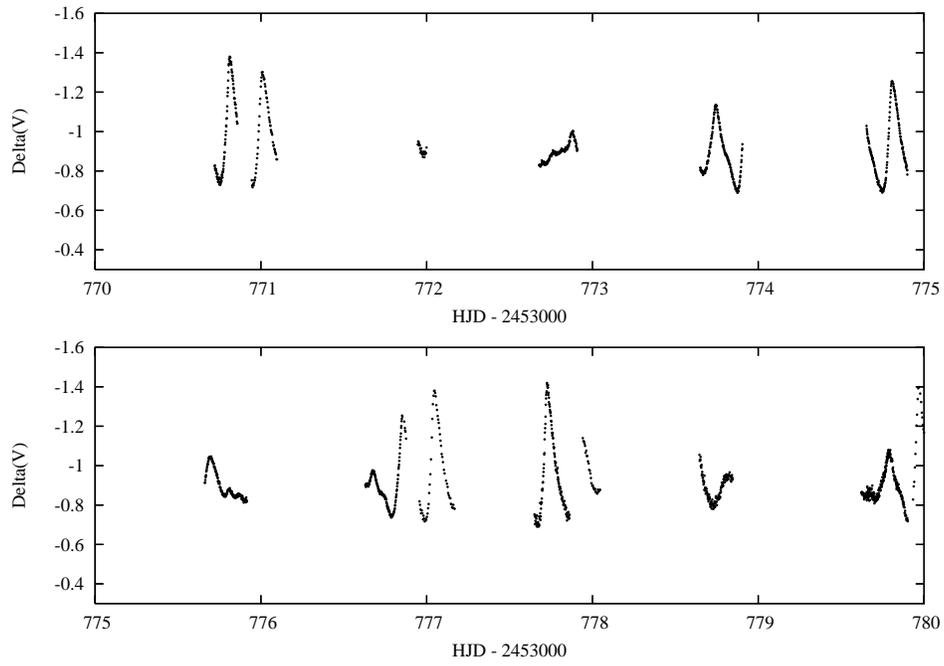}
\caption{Ten days of coverage for VX Hya.  The average photometric error
per point is less than 10 mmag, and often 5 mmag or less.  Data were
collected by four observers in the United States and one in Australia;
at least two days have more than 16 hours of photometry.}
\label{fig02}
\end{center}
\end{figure}

Global coverage was obtained for a few nights during the 2006 season,
with observers in both the US and Australia observing sequentially. 
Observers in the US typically did not overlap in coverage, but in cases
where two sites observed at the same time the data with higher
signal-to-noise was retained rather than averaging the two observers. 
The data for the 2006 season span 89 days, while those for 2007 span 122
days.  More data were collected during 2006 with all five observers
contributing observations during that season, and three contributing in
2007. The full light curve of VX Hya is shown in Figure \ref{fig01}, and
the data from HJD 2453770 to 780 are shown in Figure \ref{fig02} to
highlight the data quality.

The light curve is clearly complex in nature, a signature of
multiperiodic variability common in delta Scuti stars.  VX Hya is a
high-amplitude pulsator ($\Delta(V) > 0.1$ mag; see \citet{Breger00} for
a discussion of the classification), and so the variations are almost
certainly due to radial rather than non-radial pulsation.  (However, we note
that non-radial pulsation has been suggested for some high-amplitude stars; 
see \citet{LJR03} for example).  As we will
show, the two mode frequencies and their frequency ratio strongly suggest
purely radial double-mode pulsation with no other modes excited at
amplitudes above the limits of the photometry
(a few mmag).  The complexity of the light curve will be shown to arise
entirely from beating between these two modes.

\section{Analysis}

We Fourier analyzed the data using a deconvolving, iterative cleaning 
routine based upon the algorithm of \citet{Roberts87}.  We performed very
high resolution scans ($\delta\nu \sim 10^{-6}-10^{-5}$) between 0 and
60 cycles per day; the high resolution was required because the
theoretical errors on the frequencies are much smaller than the $1/T$
confusion limit.  Over-resolving the Fourier transforms implicitly
assumes there are no close frequency pairs with $\Delta f <1/T$, which
may or may not be valid but is a reasonable assumption for cases where
rotational splitting of nonradial modes is not expected.  For the
Fourier analysis we also used a very small gain (0.05) with a large
number of cleaning steps (12000) to keep the cleaning process as
numerically smooth as possible. 

We performed three different scans: one on the 2006 data alone, one on
the 2007 data alone, and one on the combined 2006-2007 data set. After
the spectra were computed, we then extracted the 200 highest-amplitude
peaks from each spectrum, with 200 being an arbitrarily large number
sufficient to detect all significant peaks and reach the noise level at
high frequencies.  To assess which if any of the resulting peaks were real 
we used the following selection criteria.  First, we assumed that the two 
highest-amplitude peaks $f_{0}$ and $f_{1}$ were the two known pulsation
modes.  Next, any subsequent measured peak matching a beat frequency 
calculated using $f_{\rm beat} = if_{0} + jf_{1}$, $-10\leq i,j\leq +10$ is
considered real but only if it is significantly above the local noise
level, it is present in Fourier analyses of each season's data analyzed
by itself, and that the difference between the observed beat frequency
and the predicted one is less than the statistical error on the observed
frequency.  Third, any additional peaks not matching a predicted beat frequency
must also be significantly above the local noise level.
For the noise limits, we did not calculate error bars from
the spectra based on the local noise power, but instead assumed a global
value based upon the limits of the data, and it is clear that the noise
is frequency-dependent.  At a frequency of 0 cycles per day, the
3$\sigma$ noise level is around 2-3 mmag, while it is less than 1 mmag
at 40 cycles per day.  However, using a global mean frequency did not
significantly affect the results.

\begin{figure}
\begin{center}
\includegraphics[]{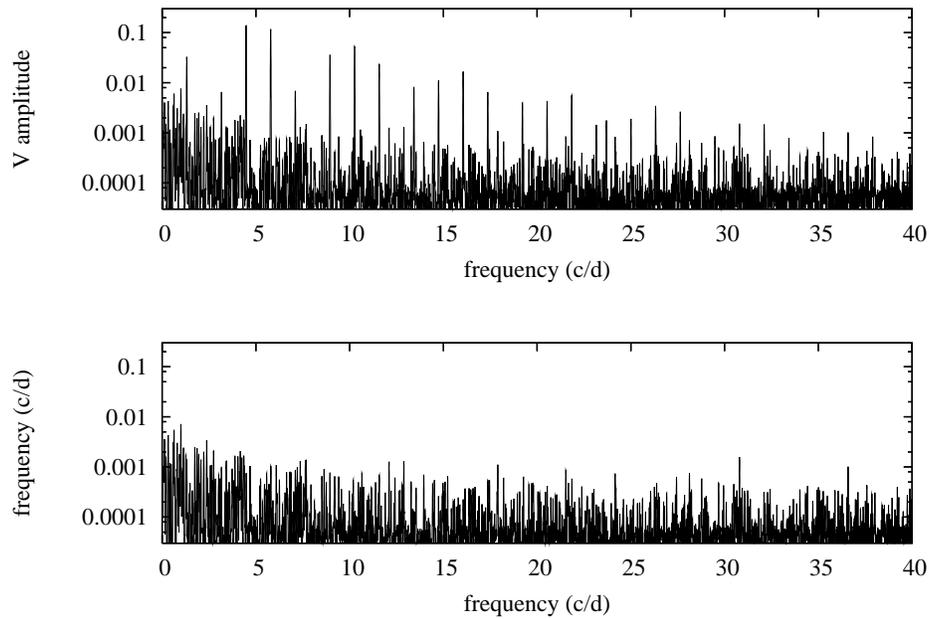}
\caption{Original Fourier spectrum (top) and prewhitened spectrum (bottom) of 
the 2006 data set.  Note that the amplitude scale is logarithmic to clearly
show the significance of the low-amplitude peaks.  Thirty frequencies were
detected in the 2006 data set having amplitude significantly above the noise
level.  Of these 30, 25 are also observed in the 2007 data set.  The 
prewhitened spectrum shows no other significant peaks, indicating that all 
of the periodic variations in VX Hya are due to the two radial modes and 
their linear combination frequencies.}
\label{fig03}
\end{center}
\end{figure}

\begin{figure}
\begin{center}
\includegraphics[]{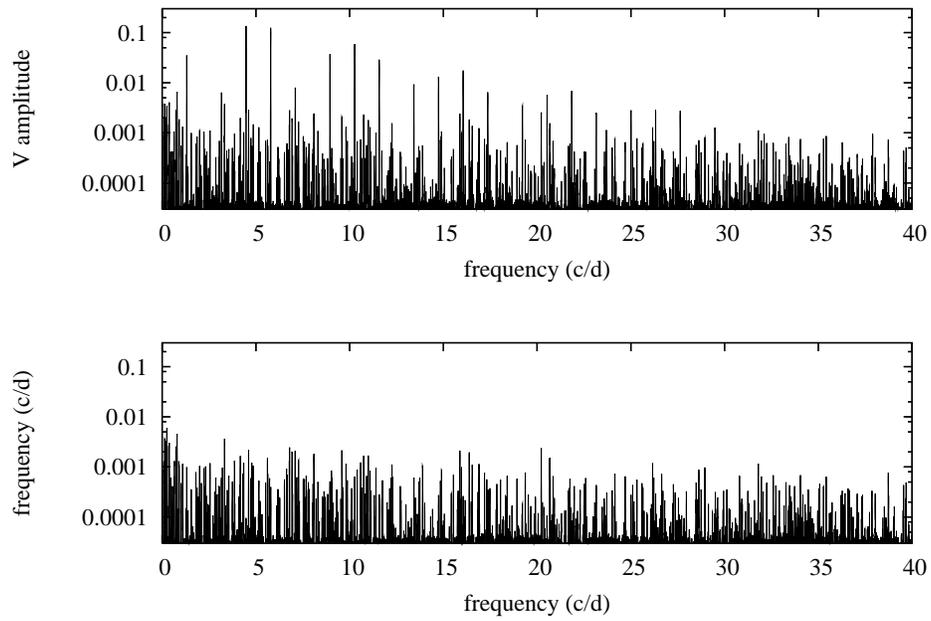}
\caption{Original Fourier spectrum (top) and prewhitened spectrum (bottom) of 
the 2007 data set.  Twenty-six frequencies were detected in the 2007 data set
having amplitude significantly above the noise level.  Of these 26, 25 are 
also observed in the 2006 data set.  As with the 2006 data, no other 
significant peaks are observed following prewhitening.}
\label{fig04}
\end{center}
\end{figure}

In Figures \ref{fig03} and \ref{fig04} we show the Fourier spectra for the 
2006 and 2007 data sets, along with the spectra calculated following 
prewhitening with the initial set of frequencies selected by the criteria
above. Twenty-five frequencies were found independently in both the
2006 and 2007 data sets; their presence in each independent data
set strongly suggests they are real.  Five additional frequencies were
unique to the 2006 data set, and one additional frequency was unique to the
2007 data set, all of which had Fourier amplitudes below 1.5 mmag. 
These peaks are questionable, but their frequencies match predicted Fourier
harmonics or beat frequencies and are above the local noise level,
and thus may be real.  Twenty-two of the 25 frequencies also appear in
the Fourier spectrum of the combined 2006-2007 data set. 

At low frequencies, there appear to be
peaks at 1 c/d and below which may be due to nightly extinction or sky
brightness variations, or to observer-to-observer differences in system
sensitivity. There are several additional peaks at frequencies below 15
c/d at the level of 1-2 mmag, all near the local (frequency-dependent)
noise level.  There are additional frequencies above 25-30 c/d which
may be additional beat frequencies, but their amplitudes are extremely
low, and are at the 1$\sigma$ uncertainty limit; they may well be real,
but we opted not to include them among the confirmed peaks.  No other
statistically significant frequencies between 0 and 60 c/d were detected
in the prewhitened spectra, suggesting that no other pulsation modes are
present in these data with measurable amplitudes. In Table \ref{table2},
we show the frequencies amplitudes, phases, and their identifications as
derived from each of the two seasons' light curves. The 1$\sigma$ error
bars on frequency, amplitude, and phase were calculated according to
\citet{LenzBreger05}.

\section{Discussion}

\begin{figure}
\begin{center}
\includegraphics[]{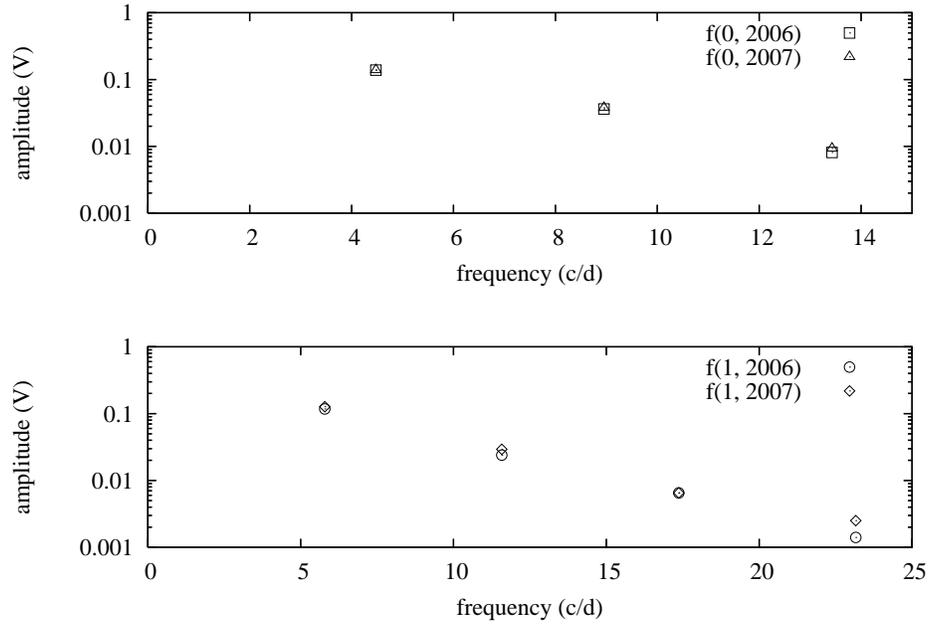}
\caption{Fourier harmonic amplitudes for $f_{0}$ (top) and $f_{1}$ (bottom)
showing the exponential decline in harmonic amplitude with increasing
harmonic order (an exponential decline appears linear in a log-linear graph).
Similar behavior is seen in other high-amplitude
pulsating stars \citep{Jurcsik05,Chadid06}.  We are therefore confident
that these peaks are legitimate detections and are correctly identified
as Fourier harmonics.}
\label{fig05}
\end{center}
\end{figure}

There are several interesting things to note about these sets of
frequencies.  First, to reiterate, all of the peaks above are (a) well
above the local noise limit, and (b) are identified as either Fourier
harmonics of one of the pulsation modes, or are beat frequencies; there
are no other statistically significant peaks in the spectrum.  Thus we
can conclude that VX Hya is only pulsating with two modes, at least at the
present epoch.  We are secure in the identification of each
peak as a harmonic or beat, as they match periods calculated using
integer combinations of the two mode frequencies to well within the
frequency errors in each case.  As further evidence of their reality, we
show the harmonic amplitudes $if(0)$ and $jf(1)$ in Figure \ref{fig05}.
The harmonic amplitudes for a given high-amplitude mode decrease
exponentially with increasing harmonic order (see \citet{Jurcsik05}), and
it is clear from the log-linear graph that these amplitudes do indeed follow
an exponential decay down to their detection limits.  The amplitudes of
beat frequencies having fixed $i$ or $j$ values exhibit the same
exponential decay.

There is no evidence for additional pulsation modes with observable
amplitudes in the VX Hya light curve, suggesting that VX Hya is a pure
double-mode pulsator.  The observed ratio of the two primary mode
frequencies $f_{0}/f_{1}$ is 0.77316, which is typical of the ratio of
the radial fundamental and first-overtone mode frequencies in high-amplitude
$\delta$ Scuti stars.
\citet{Templeton2000} showed that this ratio is consistent with a
$\delta$ Scuti star of near-solar abundances, having a mass $> 2.4
M_{\odot}$.  The pulsation frequencies
shown in Table \ref{table2} are consistent with the results of
\citet{Fitch1966}, who found that linear combinations of $f_{0}$ and
$f_{1}$ up to 4th order were sufficient to model the light curve, with
the exception of an unexplained long-term variation.
Based upon the bulk of the pulsation behavior, VX Hya appears to be a
normal high-amplitude, double-mode $\delta$ Scuti star.

Second, we found no evidence for secular changes in frequencies from the
Fourier analysis, but found strong evidence for a change in the
amplitude of $f(1)$ from 2006 to 2007.  The frequencies of each matched
peak do not differ by more than $3\sigma$ from 2006 to 2007, and so the
frequencies are constant to within the measurable errors.  (We note that
we have not yet analyzed the time-of-maximum information for these data
which is more sensitive to period changes, and we encourage a careful
$O-C$ analysis using these data.) We have, however, detected a highly
significant increase in amplitude of $f_{1}$ between 2006 and 2007.  The
amplitude in 2006 was $\Delta(V)=0.1176\pm4$, and in 2007 was
$\Delta(V)=0.1272\pm7$.  This is a change of nearly 0.01 magnitude
within 1 year.  Such a rate of change maintained over long timescales
would result in radically different pulsation spectra over time, making
it difficult to measure or model the light curve with certainty.  No
recent photometry has been published but the amplitudes obtained by
Fitch in the 1960's were substantially different.  Based upon this
two-year set of observations alone, it appears that the amplitude of at
least one of the pulsation modes of VX Hya is highly variable, and its
variability would also impact the amplitudes of its Fourier harmonics
and beat frequencies.

The sole question for VX Hya remains the long-term stability of its 
pulsation spectrum, and the origin of amplitude variation.  
\citet{Fitch1966} showed that
between 1955 and 1959, the best-fit amplitudes of the observed modes,
harmonics and beat frequencies all vary, including $f_{0}$.  The amplitudes
of all modes detected in our data lie within the range of yearly mean
amplitudes detected by Fitch, suggesting VX Hya's pulsation behavior
in the present epoch is relatively unchanged since the 1950s.  Even if
the individual component amplitudes are highly variable, they appear to
remain within a well-constrained range.  Fitch's yearly mean amplitudes
from 1955 to 1959 appear to show that the amplitude of $f_{1}$ varies 
over a larger range than does $f_{0}$.  The 450-day time span of our 
observations is too short to state whether the amplitude of $f_{0}$ is 
genuinely constant in the long term; amplitude changes in single modes have
been observed in other stars, but further observations may show that
it too is changing over the long term.  

It is possible that the temporal variations in VX Hya's behavior are
genuinely cyclical in nature as \citet{Fitch1966} implied with the 
invocation of a 4400-day supercycle, and that the present set of
observations are simply not sufficient to constrain the length and
nature of this behavior.  In principle, we could detect evidence of
a cycle having a period of up to 225 days with our data, but we
do not find any such period.  Therefore, any cyclical variations must
have longer periods than this.  

We encourage future monitoring of VX Hya, including season-long
campaigns that can investigate the changes in its pulsation
behavior that may occur on long timescales.  As we have shown, such 
campaigns can easily be done with small telescopes, and we encourage the
organization of and participation in such campaigns by the Amateur,
Professional, and Science Educator communities. 


{}

\begin{deluxetable}{cccccc}
\tabletypesize{\small}
\tablewidth{0pt}
\tablecaption{\textsc{Observer table}}
\tablehead{
\colhead{Observer} & \colhead{location} & \colhead{time zone} &
\colhead{number of obs} & \colhead{number of obs} & \colhead{telescope} \\
\colhead{} & \colhead{} & \colhead{} & \colhead{2006} & \colhead{2007}
& \colhead{}
}
\startdata
N. Butterworth & Townsville, QLD & UTC+10 & 726 & -- & 0.2-m, SBIG ST7e\\
S. Dvorak & Clermont, FL & UTC-5 & 3219 & 225 & 0.25-m, SBIG ST9XE\\
H.S. Gerner & New Berlin, WI & UTC-6 & 457 & -- & 0.25-m, SBIG ST9E\\
R. Poklar & Tucson, AZ & UTC-7 & 836 & 1203 & 0.2-m, SBIG ST9E\\
G. Samolyk & New Berlin, WI & UTC-6 & 408 & 953 & 0.25-m, SBIG ST9E\\
G. Samolyk & Big Bend, TX & UTC-6 & -- & 616 & 0.25-m, SBIG ST9XE\\
Annual Total & & & 5646 & 2997 & \\
\enddata
\tablecomments{Annual totals include all observations submitted; approximately 150 observations were not used in the final light curve.  The ``2007'' data set actually began on 2006 December 20 UT.}
\label{table1}
\end{deluxetable}

\begin{deluxetable}{cccccccc}
\tabletypesize{\tiny}
\tablecaption{Frequencies, phases, and amplitudes for the 25 peaks common to
both the 2006 and 2007 data sets.  Phases are relative to the temporal centers
of each season's data: $t_{0,2006}$ = 2453807.3515; $t_{0,2007}$ = 2454151.7757.  Numbers in
parentheses are the 1$\sigma$ errors in the last decimal place of the given
number.}
\tablehead{
\colhead{freq. (2006)} &
\colhead{$V$ amp. (2006)} &
\colhead{phase (2006)} &
\colhead{freq. (2007)} &
\colhead{$V$ amp. (2007)} &
\colhead{phase (2007)} &
\colhead{$\Delta(A)_{2007-2006}$} &
\colhead{ident.} \\
\colhead{(c/d)} &
\colhead{(mag)} &
\colhead{(radians)} &
\colhead{(c/d)} &
\colhead{(mag)} &
\colhead{(radians)} &
\colhead{(mag)} &
\colhead{$i,j$}
}
\startdata
$4.47650(2)$  & $0.1384(4)$ & $0.137(5)$   & $4.47658(2)$  & $0.1391(7)$ & $-1.0961(8)$  & $0.0007(6)$ &  $+1,+0$ \\
$5.78981(2)$  & $0.1176(4)$ & $-0.549(5)$  & $5.78983(3)$  & $0.1271(7)$ & $0.5032(9)$   & $0.0095(6)$ &  $+0,+1$ \\
$10.26629(5)$ & $0.0531(4)$ & $-2.673(1)$  & $10.26644(6)$ & $0.0591(7)$ & $-2.790(2)$   & $0.0060(6)$ &  $+1,+1$ \\
$8.95295(7)$  & $0.0363(4)$ & $-1.917(2)$  & $8.95295(9)$  & $0.0378(7)$ & $1.935(3)$    & $0.0015(6)$ &  $+2,+0$ \\
$1.31386(8)$  & $0.0328(4)$ & $2.439(2)$   & $1.31309(9)$  & $0.0352(7)$ & $-1.659(3)$   & $0.0024(6)$ &  $-1,+1$ \\
$11.5795(1)$  & $0.0239(4)$ & $-2.976(3)$  & $11.5798(1)$  & $0.0293(7)$ & $-0.918(4)$   & $0.0054(6)$ &  $+0,+2$ \\
$16.0561(1)$  & $0.0168(4)$ & $0.522(4)$   & $16.0564(2)$  & $0.0177(7)$ & $1.557(6)$    & $0.0009(6)$ &  $+1,+2$ \\
$14.7428(2)$  & $0.0110(4)$ & $1.852(6)$   & $14.7429(2)$  & $0.0134(7)$ & $0.388(9)$    & $0.0024(6)$ &  $+2,+1$ \\
$13.4293(3)$  & $0.0081(4)$ & $2.389(8)$   & $13.4296(3)$  & $0.0093(7)$ & $-1.24(1)$    & $0.0012(6)$ &  $+3,+0$ \\
$7.1029(4)$   & $0.0069(4)$ & $-0.969(9)$  & $7.1017(4)$   & $0.0080(7)$ & $2.47(1)$     & $0.0011(6)$ &  $-1,+2$ \\
$3.1641(4)$   & $0.0066(4)$ & $1.687(9)$   & $3.1641(5)$   & $0.0066(7)$ & $-2.06(2)$    & $0.0000(6)$ &  $+2,-1$ \\
$17.3699(4)$  & $0.0065(4)$ & $0.82(1)$    & $17.3692(5)$  & $0.0065(7)$ & $-2.61(2)$    & $0.0000(6)$ &  $+0,+3$ \\
$21.8448(4)$  & $0.0056(4)$ & $-2.11(1)$   & $21.8466(5)$  & $0.0071(7)$ & $0.02(2)$     & $0.0015(6)$ &  $+1,+3$ \\
$20.5330(6)$  & $0.0043(4)$ & $-1.19(1)$   & $20.533(6)$   & $0.0057(7)$ & $-1.47(2)$    & $0.0014(6)$ &  $+2,+2$ \\
$19.2196(6)$  & $0.0040(4)$ & $-0.02(2)$   & $19.2191(9)$  & $0.0036(7)$ & $-2.82(3)$    & $-0.0004(6)$ &  $+3,+1$ \\
$26.3216(7)$  & $0.0035(4)$ & $1.76(2)$    & $26.324(1)$   & $0.0029(7)$ & $2.63(4)$     & $-0.0006(6)$ &  $+2,+3$ \\
$27.6368(9)$  & $0.0027(4)$ & $1.54(2)$    & $27.635(1)$   & $0.0028(7)$ & $-1.65(4)$    & $0.0001(6)$ &  $+1,+4$ \\
$25.009(1)$   & $0.0019(4)$ & $2.95(3)$    & $25.010(1)$   & $0.0029(7)$ & $1.31(4)$     & $0.0010(6)$ &  $+3,+2$ \\
$23.697(1)$   & $0.0018(4)$ & $-1.99(3)$   & $23.695(3)$   & $0.0011(7)$ & $0.0(1)$      & $-0.0007(6)$ &  $+4,+1$ \\
$32.115(2)$   & $0.0015(4)$ & $-0.55(4)$   & $32.111(3)$   & $0.0010(7)$ & $1.1(1)$      & $-0.0005(6)$ &  $+2,+4$ \\
$23.162(2)$   & $0.0014(4)$ & $-1.63(4)$   & $23.159(1)$   & $0.0025(7)$ & $2.20(5)$     & $0.0011(6)$ &  $+0,+4$ \\
$35.279(2)$   & $0.0010(4)$ & $-2.06(6)$   & $35.277(4)$   & $0.0008(7)$ & $2.2(1)$      & $-0.0002(6)$ &  $+4,+3$ \\
$29.485(3)$   & $0.0009(4)$ & $0.23(7)$    & $29.486(3)$   & $0.0013(7)$ & $-2.00(9)$    & $0.0004(6)$ &  $+4,+2$ \\
$37.902(3)$   & $0.0008(4)$ & $2.70(8)$    & $37.900(3)$   & $0.0010(7)$ & $-0.2(1)$     & $0.0002(6)$ &  $+2,+5$ \\
$33.432(3)$   & $0.0008(4)$ & $-0.12(8)$   & $33.423(4)$   & $0.0008(7)$ & $3.1(1)$      & $0.0000(6)$ &  $+1,+5$
\enddata
\label{table2}
\end{deluxetable}

\end{document}